\documentclass{appolb}

\begin{document}

\title{The motion of test bodies with microstructure in gauge gravity models
%
%
\thanks{Presented at: {\it Myron Mathisson: his life, work, and influence on current research}, Stefan Banach International Mathematical Center, Warsaw, Poland, 18 -- 20 October, 2007}
}
\author{Dirk Puetzfeld
\address{Institute of Theoretical Astrophysics, University of Oslo, P.O. Box 1029, 0315 Oslo, Norway}
}

\maketitle

\begin{abstract}
We report on the explicit form of the equations of motion of pole-dipole particles for a very large class of gravitational theories. The non-Riemannian framework in which the equations are derived allows for a unified description of nearly all known gravitational theories. The propagation equations are obtained with the help of a multipole expansion method from the conservation laws that follow from Noether's theorem. The well-known propagation equations of general relativity, e.g., as given by Mathisson and Papapetrou, represent a special case in our general framework. Our formalism allows for a direct identification of the couplings between the matter currents and the gravitational field strengths in gauge gravity models. In particular, it illustrates the need for matter with microstructure for the detection of non-Riemannian spacetime geometries.
\end{abstract}

\section{Introduction.}\label{introduction_sec}

The link between the field equations and the equations of motion in gravitational theories has been subject to many works. In the context of the theory of general relativity the earliest accounts of this feature can be found in the works of Weyl \cite{Weyl:1923}, Eddington \cite{Eddington:1924}, Einstein \& Grommer \cite{EinsteinGrommer:1927}, Einstein, Infeld \& Hoffmann \cite{EinsteinInfeldHoffman:1938}, as well as Fock \cite{Fock:1939}. Nowadays this is customarily addressed as the ``problem of motion''.

One of the early contributers in this field - to whom's life and work this conference is dedicated - was Myron Mathisson, who published a series of works \cite{Mathisson:1931:3,Mathisson:1931:1,Mathisson:1931:2} on the problem of motion, in particular the systematic account \cite{Mathisson:1937}, over seventy years ago. 

Here we provide an answer to the question of how test particles move under the influence of the gravitational field in gauge gravity models. We base our considerations on the theory of metric-affine gravity, see \cite{Hehl:1995} for a review. Metric-affine gravity provides a proper physical and mathematical framework for many gravitational models, in particular it allows for a unified description of a large class of alternative gravity theories. In contrast to the theory of general relativity, the spacetime must no longer be structured according to the Riemannian scheme in such a theory. Furthermore, not only the energy-momentum, but also the hypermomentum -- which describes intrinsic properties of matter like the spin, dilation, and shear currents -- may act as source of the gravitational field. 

To our knowledge the program laid out by Mathisson \cite{Mathisson:1937}, and its later realization by Dixon \cite{Dixon:1970:1,Dixon:1970:2,Dixon:1974}, remains unrivaled with respect to its mathematical rigor. Nevertheless, we do not employ it here, but make use of a multipole approximation scheme in the spirit of Papapetrou \cite{Papapetrou:1951:3} to derive the equations of motion of pole-dipole test particles. The reason for this is the simplicity of the scheme and its direct applicability to the type of theories which we are interested in. 

\section{Metric-affine gravity.}\label{mag_intro_sec}

In metric-affine gravity, besides the usual ``weak'' long-range Newton-Einstein type gravity, described by the metric $g_{ij}$ of spacetime, an additional ``strong'' short-range gravity piece is mediated by the independent linear connection $\Gamma_{i j}{}^{k}$. It is different from the Riemannian (Christoffel) connection, and the difference is described in terms of the tensors of nonmetricity $Q_{ijk}:=-\nabla_i g_{j k}$ and of the torsion $S_{ij}{}^{k}:=\Gamma_{i j}{}^{k}-\Gamma_{j i}{}^{k}$ which are also manifest in the non-Riemannian pieces of the curvature $R_{ijk}{}^{l}$.

The matter currents, which are the sources of the gravitational field, are obtained by variation of the matter Lagrangian with respect to the gravitational potentials (metric $g_{ij}$, coframe $h^\alpha_j$, connection $\Gamma_{i j}{}^{k}$). This yields the canonical energy-momentum $T_i{}^j:=h^\alpha_i\delta {L}_{\rm mat} / \delta h^\alpha_j$, the metrical energy-momentum $t^{ij}:=2 \delta {L}_{\rm mat} / \delta g_{ij}$, and the hypermomentum $\Delta^i{}_j{}^k := \delta { L}_{\rm mat} / \delta\Gamma_{ki}{}^j$ current.
 
\subsection{Energy-momentum conservation.}\label{em_conservation_subsec}

The Noether theorem for the diffeomorphism invariance of the matter action yields the conservation law, see \cite{Obukhov:Rubilar:2006} for a recent review, of the energy-momentum 
\begin{equation}
{\stackrel{\{\,\}}{\nabla}}_j\left(T_i{}^j - N_{ikl}\,\Delta^{klj}\right)=\big({\stackrel{\{\,\}}{R}}_{ijkl} - {\stackrel{\{\,\}}{\nabla}}_i \,N_{jkl}\big)\Delta^{klj}.\label{DT1}
\end{equation}
Here, and in the following, curled braces ``$\{ \}$'' denote objects which are based on the symmetric Riemannian connection (Christoffel symbols), and $N_{ij}{}^k :={\stackrel{\{\,\}}{\Gamma}}_{ij}{}^k-\Gamma_{ij}{}^k$ represents the so-called distorsion tensor. Equation (\ref{DT1}) can be identically rewritten as 
\begin{equation}
{\stackrel{\{\,\}}{\nabla}}_j\,T_i{}^j = \widehat{R}_{ijkl}\,\Delta^{klj} + N_{ikl}\,{\stackrel{\{\,\}}{\nabla}}_j\Delta^{klj},\label{DT2}
\end{equation}
where we introduced $\widehat{R}_{ijkl} := {\stackrel{\{\,\}}{R}}_{ijkl} - {\stackrel{\{\,\}}{\nabla}}_i N_{jkl} + {\stackrel{\{\,\}}{\nabla}}_j N_{ikl}$.

\subsection{Hypermomentum conservation}\label{hm_conservation_subsec}

The Noether theorem for the invariance of metric-affine gravity under the local general linear group yields (on the ``mass shell", i.e., when the matter satisfies the field equations): 
\begin{equation}
{\stackrel{\{\,\}}{\nabla}}_j\,\Delta^{klj} - N_{ij}{}^k\Delta^{jli} + N^{jli}\Delta^k{}_{ij} + T^{lk} - t^{kl} = 0.\label{GLc}
\end{equation}

\section{Equations of motion.}\label{sec_propa_eq_ys}

Denoting the densities of objects by a tilde ``$\widetilde{{\phantom{A}}}$'' the conservation equations for the canonical energy-momentum current (\ref{DT2}) and hypermomentum current (\ref{GLc}), take the following form  
\begin{eqnarray}
\partial_j\widetilde{T}{}_{i}{}^{j} &=&R_{ijk}{}^{l} \widetilde{\Delta }^{k}{}_{l}{}^{j}+\Gamma _{ij}{}^{k} \widetilde{T}{}_{k}{}^{j}+N_{ij}{}^{k}\widetilde{t}^{j}{}_{k}, \label{em_conservation_ala_DY_1} \\
\partial_j\widetilde{\Delta }^{k}{}_{l}{}^{j} &=&\Gamma _{jl}{}^{m} \widetilde{\Delta }^{k}{}_{m}{}^{j}-\Gamma _{mj}{}^{k} \widetilde{\Delta }^{j}{}_{l}{}^{m}-\widetilde{T}{}_{l}{}^{k}+\widetilde{t}^{k}{}_{l}. \label{em_conservation_ala_DY_2}
\end{eqnarray}
Note that $\Gamma_{ij}{}^{k}$ represents the full connection. The last two equations should be compared to (42) and (43) in \cite{Stoeger:Yasskin:1980}, as well as (3.13) and (3.14) in \cite{Nomura:etal:1991}. 

\subsection{Integrated moments.}

We introduce the integrated multipole moments of the matter currents as follows:
\begin{eqnarray}
\underline{\Delta}^{b_{1}\cdots b_{n}i}{}_{j}{}^{k} &:&=\int \left( \prod\limits_{\alpha=1}^{n}\delta x^{b_{\alpha }}\right) \widetilde{\Delta }^{i}{}_{j}{}^{k}, \nonumber \\
\underline{T}^{b_{1}\cdots b_{n}}{}_{i}{}^{j} &:&=\int \left( \prod\limits_{\alpha =1}^{n}\delta x^{b_{\alpha }}\right) \widetilde{T}_{i}{}^{j},  \nonumber \\
\underline{t}^{b_{1}\cdots b_{n}i}{}_{j} &:&=\int \left( \prod\limits_{\alpha=1}^{n}\delta x^{b_{\alpha }}\right) \widetilde{t}^{i}{}_{j}. \label{DY_int_moments_definitions}
\end{eqnarray}
Here $\delta x^{a}:=x^{a}-Y^{a},$ where $Y$ parametrizes the worldline. Note the difference with respect to method of Mathisson and Dixon, who use an implicit definition of the moments, at this point. The integrals are taken over a 3-dimensional slice $\Sigma(t)$, at a time $t$, of the world tube of a test body. We use the condensed notation
\begin{equation}
\int\,f = \int_{\Sigma(t)}\,f(x)\,d^3x.
\end{equation} 

\subsection{Propagation equations for pole-dipole particles.}

From the expressions (\ref{em_conservation_ala_DY_1}) and (\ref{em_conservation_ala_DY_2}) we can derive the propagation equations for pole-dipole particles.\footnote{More details on the derivation can be found in \cite{Puetzfeld:Obukhov:2007}.} For such bodies the following moments are assumed to be non-vanishing: $\underline{\Delta}^{i}{}_{j}{}^{k}, \underline{T}_{i}{}^{j}, \underline{T}^{i}{}_{j}{}^{k}, \underline{t}^{i}{}_{j},$ and $\underline{t}^{ij}{}_{k}$. The expansion of geometrical quantities around the worldline $Y(t)$ of the body into a power series in $\delta x^{a},$ reads
\begin{eqnarray}
\left. R_{ijk}{}^{l} \right| _{x} &=&\left. R_{ijk}{}^{l} \right| _{Y} +\delta x^{a}\left. R_{ijk}{}^{l}{}_{,a}\right| _{Y}+\dots ,  \nonumber \\
\left. \Gamma _{ij}{}^{k}\right| _{x} &=&\left. \Gamma _{ij}{}^{k}\right| _{Y}+\delta x^{a}\left. \Gamma _{ij}{}^{k}{}_{,a}\right| _{Y}+\dots , \nonumber \\
\left. N_{ij}{}^{k}\right| _{x} &=&\left. N_{ij}{}^{k}\right| _{Y}+\delta x^{a} \left. N_{ij}{}^{k}{}_{,a}\right| _{Y}+\dots . \label{DY_geom_quant_taylor}
\end{eqnarray}
In the following we are going to suppress the dependencies on the points at which certain quantities are evaluated. Furthermore, we introduce new names for the integrated quantities as follows: $\underline{P}_i := \underline{T}_i{}^0$ denotes the integrated 4-momentum, $\underline{L}^k{}_l := \underline{T}^k{}_l{}^0$ the total orbital canonical energy-momentum, $\underline{Y}^k{}_l := \underline{\Delta}^k{}_l{}^0$ the integrated intrinsic hypermomentum, and 
\begin{equation}
{\cal P}_{i}:=\underline{P}_{i}-N_{ik}{}^{l}\underline{Y}^{k}{}_{l} -{\stackrel{\{\,\}}{\Gamma}}_{ik}{}^{l}\underline{L}^{k}{}_{l}{}, \label{Ptot}
\end{equation}
the generalized total 4-momentum. In addition, we introduce a shorter notation for the ``convective currents": For the intrinsic hypermomentum we have ${\stackrel{(c)}{\underline{\Delta}}}{}^k{}_l{}^m := \underline{\Delta}^k{}_l{}^m - v^m\,\underline{\Delta}^k{}_l{}^0$, and for the orbital canonical energy-momentum ${\stackrel {(c)}{\underline{T}}}{}^k{}_l{}^m :=\underline{T}^k{}_l{}^m - v^m\,\underline{T}^k{}_l{}^0$. The fluid derivative is defined as follows $\nabla_v\,\underline{Y}^i{}_k:=d/dt \, \underline{Y}^i{}_k + v^m \Gamma_{mj}{}^{i} \underline{Y}^j{}_k - v^m \Gamma_{mk}{}^{j} \underline{Y}^i{}_j$, here $v^a:=dY^a/dt$. With this notation, the integrated version of the conservation laws (\ref{em_conservation_ala_DY_1}) and (\ref{em_conservation_ala_DY_2}) yields the propagation equations
\begin{eqnarray}
{\stackrel{\{\,\}}{\nabla}}_{v}{\cal P}_{i}&=&\left({\stackrel{\{\,\}}{R}}_{ijk}{}^{l}-{\stackrel{\{\,\}}{\nabla}}_{i} N_{jk}{}^{l}\right)\underline{\Delta}^{k}{}_{l}{}^{j}+{\stackrel{\{\,\}}{R}}_{ijk}{}^{l} {\stackrel{(c)}{\underline{T}}}{}^{k}{}_{l}{}^{j}+{\stackrel{\{\,\}}{R}}_{kji}{}^{l} {\underline{L}}{}^{k}{}_{l} v^j,\label{DY_pd_prop_eq_1a} \\
\underline{T}_k{}^i &=& v^i\,\underline{P}_k + {\frac {d}{dt}} \,\underline{L}^i{}_k - {\stackrel{\{\,\}}{\Gamma}}_{kj}{}^{l} \, \underline{T}^i{}_l{}^j + N_{kj}{}^{l}\,{\stackrel {(c)}{\underline{\Delta}}}{}^j{}_l{}^i,\label{DY_pd_prop_eq_2a}\\
{\stackrel {(c)}{\underline{T}}}{}^{(a}{}_i{}^{b)} &=& 0, \label{DY_pd_prop_eq_3a}\\
\nabla_v\,\underline{Y}^i{}_k &=& -\,\underline{T}_k{}^i + \underline{t}^i{}_k - \Gamma_{jl}{}^{i}\,{\stackrel {(c)}{\underline{\Delta}}} {}^l{}_k{}^j +\Gamma_{jk}{}^{l}\,{\stackrel {(c)}{\underline{\Delta}}}{}^i {}_l{}^j, \label{DY_pd_prop_eq_4a}\\
{\stackrel {(c)}{\underline{\Delta}}}{}^k{}_l{}^a &=& \underline{T}^a{}_l{}^k - \underline{t}^{ak}{}_l.\label{DY_pd_prop_eq_5a}
\end{eqnarray}
The propagation equation (\ref{DY_pd_prop_eq_1a}) for the generalized total 4-momentum should be compared to (5.7) in \cite{Papapetrou:1951:3}, (53) in \cite{Stoeger:Yasskin:1980}, and (5.19) in \cite{Nomura:etal:1991}. Equation (\ref{DY_pd_prop_eq_2a}) describes the canonical energy-momentum in terms of the usual combination of the ``translational" plus ``orbital" contributions (the first two terms), plus the additional contribution of the first moments. Equation (\ref{DY_pd_prop_eq_3a}) simply tells us that the convective current ${\stackrel{(c)}{\underline{T}}}{}^{a}{}_i{}^{b}$ is antisymmetric in the upper indices $a$ and $b$. The next equation (\ref{DY_pd_prop_eq_4a}) is actually an equation of motion for the intrinsic hypermomentum. Its  form closely follows the Noether conservation law of the hypermomentum, cf.\ (\ref{GLc}). Finally, the equation (\ref{DY_pd_prop_eq_5a}) expresses the convective intrinsic hypermomentum current in terms of the first moments of the energy-momentum. 

\section{Conclusions.}

The set (\ref{DY_pd_prop_eq_1a})-(\ref{DY_pd_prop_eq_5a}) can be viewed as the generalization of the well-known Mathisson-Papapetrou equations for pole-dipole test particles to almost any known gauge gravity model. The Mathisson-Papapetrou equations, as well as previous results in the context of spacetimes with torsion \cite{Stoeger:Yasskin:1980,Nomura:etal:1991}, can be easily recovered in our framework, see \cite{Puetzfeld:Obukhov:2007} for details. 

We notice a general feature that characterizes the coupling between the physical objects (currents) with the geometrical objects (metric, connection, and the derived quantities). Namely, the {\it intrinsic} current (the one that is truly {\it microscopic}, which arises from the averaging over the medium with the elements with microstructure, i.e., that possess internal degrees of freedom) couples to the {\it non-Riemannian} geometric quantities, see the second term on the r.h.s.\ of (\ref{Ptot}) and the first term on the r.h.s.\ of (\ref{DY_pd_prop_eq_1a}). In contrast to this, the {\it orbital} canonical energy-momentum (which is induced by the {\it macroscopic} dynamics of the rotating and deformable body) is only coupled to the purely Riemannian  geometric variables and never couples to the non-Riemannian geometry, see the last terms on the right-hand sides of (\ref{Ptot}) and (\ref{DY_pd_prop_eq_1a}). This observation demonstrates that the possible presence of the non-Riemannian geometry (in particular, of torsion and nonmetricity) can only be tested with the help of bodies that are constructed from media with microstructure (spin, dilaton charge, and intrinsic shear). Test particles, composed from usual matter {\it without} microstructure, are {\it not} affected by the non-Riemannian geometry, and they thus {\it cannot} be used for the detection of the torsion and the nonmetricity. 

It would be very interesting to carry out an analysis in the spirit of Mathisson and to compare the results to the ones obtained in this work.

\section*{Acknowledgments}
I am very grateful to A.\ Trautman (Univ.\ Warsaw) and the organizers for the invitation to Warsaw and their kind support. Furthermore, I am greatly indebted to Y.N.\ Obukhov (Univ.\ Cologne \& Moscow State Univ.) and F.W.\ Hehl (Univ.\ Cologne \& Univ.\ Missouri-Columbia) for many stimulating discussions as well as their constant advice and interest in the subject. This work was supported by the Research Council of Norway under the project number 162830. 

\bibliographystyle{abbrv}
\bibliography{The_motion_of_test_bodies_with_microstructure_in_gauge_gravity_models}

\end{document}